\documentclass[useAMS,usenatbib,a4paper]{mn2e}

\usepackage{aas_macros}

\usepackage{graphicx}
\usepackage{mathptmx}
\usepackage{amsmath}
\usepackage{color}

\newcommand{\<}{\begin{eqnarray}}
\renewcommand{\>}{\end{eqnarray}} 
\renewcommand{\bar}{\overline}
\renewcommand{\tilde}{\widetilde}
\renewcommand{\hat}{\widehat}
\newcommand{\Msun}{\ensuremath{\rmn{M}_\odot}}

\newlength{\halfcolumn}
\newlength{\fullcolumn}
\newlength{\fullcolumnspace}
\title[Mass segregation in star formation simulations]{
Global mass segregation in hydrodynamical simulations of star formation
}
\author[Th. Maschberger \& C. J. Clarke]
{Th. Maschberger$^{1,2,3}$\thanks{e-mail: thomas.maschberger@obs.ujf-grenoble.fr}
\& 
C.J. Clarke$^3$
\\
\small \it 
$^1$ Institut de Plan{\'e}tologie et d{'}Astrophysique de Grenoble, BP 53, F-38041 Grenoble C{\'e}dex 9, France\\
$^2$ Argelander-Institut f{\"u}r Astronomie, Auf dem H{\"u}gel 71, D-53121 Bonn, Germany\\
$^3$ Institute of Astronomy, Madingley Road, Cambridge CB3 0HA, England
}
\date{MNRAS accepted}

\pagerange{\pageref{firstpage}--\pageref{lastpage}} \pubyear{2011}

\begin{document}
\label{firstpage}


\maketitle

\begin{abstract}
Recent analyses of mass segregation diagnostics in star forming regions
invite a comparison with the output of hydrodynamic simulations of
star formation. 
In this work we investigate the state of mass segregation of `stars'
(i.e.  sink particles in the simulations) in the case of hydrodynamical simulations which omit  feedback.
We first discuss methods to quantify mass segregation in substructured regions, either based on the minimum spanning tree (Allison's $\Lambda$), or through
analysis of correlations between stellar mass and local  stellar 
surface number densities.
We find that  the presence of even a  single `outlier' (i.e. a massive
object far from other stars) can cause the Allison $\Lambda$ method to
describe the system as inversely mass segregated, even where in reality the
most massive sink particles are overwhelmingly in the centres of the subclusters. We
demonstrate that a variant of the $\Lambda$ method is less 
susceptible to this tendency but also argue for an alternative
representation of the data in the plane of stellar mass versus local surface 
number density. 

 The hydrodynamical simulations show global mass segregation from very
early times  which continues throughout the simulation, being only mildly influenced during
sub-cluster merging.

We find that up to $\approx$ 2--3\% of the ``massive'' sink particles ($m > 2.5\ \Msun$) are in relative isolation because they have formed there, although other sink particles can form later in their vicinity.
Ejections of massive sinks from subclusters do not contribute to the number of isolated massive sink particles, as the gravitational softening in the calculation suppresses this process.

\end{abstract}

\begin{keywords}
stars: formation -- open clusters and associations: general -- methods: data analysis
\end{keywords}

\section{Introduction}

 The relative spatial distributions of stars of different masses can
give 
valuable insights into  the evolution of young star forming regions.
Traditionally,  what is termed `mass segregation' (i.e. the concentration
of more massive stars within dense cluster cores) has been interpreted
as a stellar dynamical effect (i.e. the result of two-body relaxation), although
there are apparently cases where the system is so young that the
central concentration of massive stars must be primordial \citep{bonnell+davies1998}.
More recently,  hydrodynamical simulations of star formation
in turbulent molecular clouds have illustrated the bottom-up creation of star
clusters through sub-cluster merging, which rather blunts the distinction
between primordial and dynamical mass segregation
(see e.g. \citealp{AllisonGoodwinParker-2009} and \citealp{AllisonGoodwinParker-2010}  for purely stellar-dynamical aspects of primordial vs. dynamical mass segregation).
 What has now become
important is instead to assess the realism of such simulations by
comparing their mass segregation characteristics with those of
observed star forming regions. With this in mind it is important
to develop robust and flexible diagnostics that can be
applied to simulations and observations alike.

 Currently there are a variety of claims in the literature about the
degree of mass segregation in star forming regions:
\citet{ParkerBouvierGoodwin-2011} find {\it ``inverse''} mass segregation for Taurus, implying
that massive stars are more widely distributed than average and found preferentially isolated compared to intermediate-mass stars.
On the other hand \citet{KirkMyers-2011} find a concentration of massive stars in the centres of groups in Taurus, Lupus3, ChaI, and IC348;
\citet{schmeja-etal2008} report no signs of mass segregation for NGC1333 and L1688, but find that the older regions IC348 and Serpens are mass segregated.
Mass segregation has also been reported for more massive young star clusters like the ONC \citep{hillenbrand+hartmann1998,allison-etal2009a}, NGC3603 \citep{stolte-etal2006} or NGC2244 (\citealp{chen-etal2007}) though  \citet{2008ApJ...675..464W} find the latter not to be mass segregated.  (See
however \citealp{2009Ap&SS.324..113A} for a discussion of how incompleteness may affect these conclusions).

The association of stars with gas filaments in young star forming regions \citep[see e.g. the recent Herschel results for  Aquila][]{konyves-etal2010,bontemps-etal2010,andre-etal2010} illustrates that {\it pure} $n$-body simulations are not sufficient to describe the dynamical evolution of these regions.
Larger-scale hydrodynamical simulations of star cluster formation have been performed e.g. by \citet{bonnell-etal2003,bonnell-etal2008} or \citet{bate2009a}.
\citet{bate2009a} reported no evidence for mass segregation in
his simulations, when analysed using  cumulative radial fractions in different
stellar mass ranges, but the analysis of \citet{moeckel+bonnell2009b} 
(using convex hulls) found that the most massive stars were indeed segregated,
at a level comparable to that observed in the Orion Nebula Cluster.
In this work we analyse the global mass segregation in the simulations of \citet{bonnell-etal2003} and \citet{bonnell-etal2008}.
This extends the analysis of \citet{maschberger-etal2010}, where we have already studied mass segregation on a subcluster-by-subcluster scale.

For the analysis of mass segregation in a star forming region it is necessary to use a method that does not rely on spherical symmetry, as significant amounts of substructure can still be present.
Additionally, the method should be able to detect a concentration of only the most massive stars, of which typically only a small number exists.
\citet{allison-etal2009a} presented a method that is able to cope with both substructure and small numbers, (the $\Lambda$ measure) and 
\citet{ParkerBouvierGoodwin-2011} used this method to show that the massive stars in the Taurus star forming region are {\it inversely mass segregated}.
This is somewhat in contrast to the result of \citet{KirkMyers-2011}, who also analysed the Taurus region (albeit a different observational sample) and came to the opposite conclusion, namely mass segregation of the massive stars.
However, \citet{KirkMyers-2011} analysed individual subclusters of stars in Taurus using the radial distance of the massive stars to the geometrical centre of the subcluster.
Because of the different methods employed, it is not clear whether the \citet{KirkMyers-2011} result of local mass segregation may be directly compared to the \citet{ParkerBouvierGoodwin-2011} result of global inverse mass segregation.

 In our previous analysis of star formation simulations \citep{maschberger-etal2010} we followed both methodologies: i.e.
the approach of \citet{KirkMyers-2011}, using radial distances for lower-$n$ subclusters, and the approach of \citet{ParkerBouvierGoodwin-2011}, using the $\Lambda$ measure for sufficiently populous subclusters.
We found that generally  mass segregation prevails {\it within  subclusters} at all times apart from the very earliest stages of their  formation.
Again, our result is not directly comparable to \citet{ParkerBouvierGoodwin-2011}, because we have not analysed the simulation as a whole, but individual subclusters.

Therefore  in this work we analyse the simulations of \citet{bonnell-etal2003} and \citet{bonnell-etal2008} in the same manner as \citet{ParkerBouvierGoodwin-2011} in order to assess whether the simulations produce stellar distributions
that are compatible with those observed throughout the Taurus star forming
region.
During the analysis we found that the $\Lambda$ method can produce
misleading results when a small number of high-mass stars are present in low-density regions and that this 
might be the reason for the different results of \citet{ParkerBouvierGoodwin-2011} and \citet{KirkMyers-2011}.
In order to address this shortcoming we suggest   a modification to the
$\Lambda$ method that is less sensitive to a small number of massive
outliers. We  also point out that plots of stellar mass as a function
of surface density provide a ready way to compare the output of
simulations with observations. Such plots provide helpful visualisation
in arbitrary geometry and allow one to
easily spot and assess the presence of isolated high-mass stars.

The outline of this paper is as follows:
We first describe the hydrodynamical simulations (Section \ref{simulations}).
In the following section we describe the $\Lambda$ method to quantify mass segregation in substructured data.
Section \ref{sec_sigma} describes the state of mass segregation in the simulations when analysed in terms of local stellar densities.
We finish with a summary and discussion in Section \ref{conclusions}.

\section{Hydrodynamical simulations}\label{simulations}

The data for our analysis are from the smoothed particle hydrodynamics (SPH) simulations by \citet{bonnell-etal2003} and \citet{bonnell-etal2008}, as also analysed in \citet{maschberger-etal2010}.
In these cited works one can find detailed descriptions of the setup of the calculations and their evolution.

\citet{bonnell-etal2003} followed the self-gravitating evolution of a gas sphere containing $1000\ \Msun$ gas in a diameter of $1\ \mathrm{pc}$ with a temperature of $10\ \mathrm{K}$, which is initially marginally unbound.
An initial divergence-free random Gaussian velocity with a power spectrum $P(k) \propto k^{-4}$ models initial turbulent motions.
The gas is kept isothermal throughout the calculation and  feedback is not included.
Sink particles replace dense regions when a critical density of $1.5 \times 10^{-15}\ \mathrm{g}\ \mathrm{cm}^{-3}$ is exceeded, and can continue to accrete either if a gas particle becomes gravitationally bound within a sink radius of $200\ \mathrm{au}$ or if a gas particle moves within the accretion radius of $40\ \mathrm{au}$.
The mass resolution is $\approx 0.1\ \Msun$, and gravitational forces between sinks are smoothed at $160\ \mathrm{au}$.
In this simulation one final ``star cluster'' is formed at the end of the simulation time, at $\approx 4.8 \times 10^5\ \mathrm{yr}$ or $\approx$ 2.5 initial free-fall times, and it contains $\approx$ 560 sink particles with $m > 0.08\ \Msun$.

The simulation of \citet{bonnell-etal2008} starts with a gas mass of $10^4\ \Msun$, arranged in a cylinder having $10\ \mathrm{pc}$ length and $3\ \mathrm{pc}$ diameter with a linear density gradient along the main axis (33\ \% higher density than average on one end, 33\ \% lower on the other).
The initial turbulence is modelled as in the $10^3\ \Msun$ calculation, but the gas follows a barotropic equation of state during the simulation.
Again this simulation does not include feedback effects.
and star formation is modelled via sink particles, with a critical density of $6.8 \times 10^{-14}\ \mathrm{g}\ \mathrm{cm}^{-3}$, a sink radius of $200\ \mathrm{au}$ and an accretion radius of $40\ \mathrm{au}$.
The gravitational softening length is $40\ \mathrm{au}$.
At the end, after  $\approx 6.5 \times 10^{5}\ \mathrm{yr}$ or $\approx 1$ initial free fall time, $\approx$ 1900 sink particles are formed ($m>0.08\ \Msun$).
The spatial distribution of the sink particles is still substructured, with several larger clusters and a population along filaments.

\section{Measuring mass segregation using minimum spanning trees}\label{sec_lambda}

\subsection{Method}
\begin{figure}
\begin{center}
\includegraphics[width=0.4\textwidth]{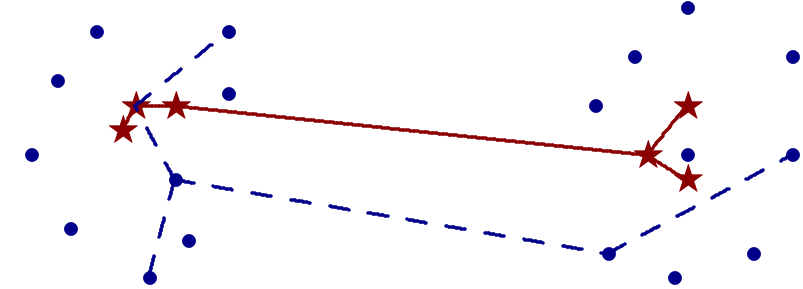}
\end{center}
\caption{\label{sketch_substructure}
Sketch illustrating the ability of MST methods to detect mass segregation in substructured regions.
Stars indicate the position of massive stars, with their MST shown as solid lines, and points the location of low-mass stars.
An MST of six random stars is shown with dashed lines.
Typically, a random set of six stars contains stars from both subclusters.
Therefore, both the MST for massive and random stars contain the long connection between the two subclusters, which averages out when they are compared.
}
\end{figure}

The basis for calculating $\Lambda$ of \citet{allison-etal2009a} is the minimum spanning tree (MST), a graph-theoretical concept, which is the unique connection of a set of points such that there are no closed loops and with the property that the total length of the connections (edges) is minimal.
Algorithms to calculate the MST, further properties, and some applications of the MST to analyse the structure in spatial data can be found in \citet{zahn1971}.
A schematic sketch illustrating how $\Lambda$ works is given in Figure \ref{sketch_substructure}, showing the spatial distribution of a mass segregated region, with dots for low-mass stars and stars for high-mass stars.
To obtain $\Lambda$, calculate first the MST of the six most massive stars of a star cluster, shown with the solid lines in Fig. \ref{sketch_substructure}.
Denote the average edge length of this MST with $\bar{l}_\mathrm{massive}$.
Consider now the MST of a set of six stars that are randomly taken from the star cluster (e.g. the dashed lines in Fig. \ref{sketch_substructure}), with average edge length $\bar{l}_\mathrm{random}$.
The MST of one particular random set of six stars contains not much information, but the sample average of the average length, $\bar{\bar{l}_\mathrm{random}}$, calculated from a number of sets containing six randomly drawn stars, characterises the spatial distribution of stars in the star cluster.
To quantify mass segregation one now compares $\bar{l}_\mathrm{massive}$ with $\bar{\bar{l}_\mathrm{random}}$.
Mass segregation is commonly understood as a more central concentration of the massive stars compared to all stars of the star cluster or, more generally, a more compact configuration of the massive stars.
Therefore, in a mass segregated cluster, the massive stars will have a much shorter average MST edge length than random stars, i.e. if $\bar{l}_\mathrm{massive} <  \bar{\bar{l}_\mathrm{random}}$ then the cluster is mass segregated.
The statistical significance of this comparison can be expressed in standard deviations of $\bar{\bar{l}_\mathrm{random}}$.

The sketch of Fig. \ref{sketch_substructure} shows a substructured region (or, rather, two subclusters), to illustrate the ability of $\Lambda$ to deal with substructure.
The massive stars reside in the centres of the subclusters.
Their MST contains four short edges (combining the stars within the subclusters) and one long edge, that connects the two subclusters.
Similarly, an MST of six random stars will contain typically stars from both subclusters, so that the random MST contains some medium-length edges plus the long connection between the subclusters.
Therefore, $\bar{l}_\mathrm{massive} $``$=$''$($short$+$connection$)/6$ is still smaller than $\bar{\bar{l}_\mathrm{random}}$``$=$''$($medium$+$connection)$/6$, even in the presence of substructure.

For the definition of the mass segregation measure we follow \citep{ParkerBouvierGoodwin-2011} by calculating  $\Lambda$  within a sequence of
different mass ranges (instead of calculating a measure for all stars from the most massive down to the i-th most massive star, as in \citet{allison-etal2009a}).
The length of $40$ stars for the moving window follows the choice of \citep{ParkerBouvierGoodwin-2011}.
We denote the Allison et al./Parker et al. measure with $\bar{\Lambda}$ to indicate that it has been calculated with the average MST length.
$\bar{\Lambda}$ is derived from the mass-sorted sample of stars ($m_{(1)} > m_{(2)}$ \dots) for the $i$-th to the $i+40$-th most massive star according to 
\< \bar{\Lambda}_{(i)} ( \bar{m_{(i)} } ) &=& \frac{ \bar{\bar{l}_{40} } } {\;\;\; \bar{l}_{(i),(i+40)} \;\;\; } \>
where $\bar{m_{(i)}}$ is the average mass of the stars from the $i$-th to the $i+40$-th most massive star, $\bar{\bar{l}_{40}}$ is the sample average MST edge length of 40 random stars ($\bar{\bar{l}_\mathrm{random}}$), and $\bar{l}_{(i),(i+40)}$ is the average edge length of the MST containing the $i$-th to the $i+40$-th most massive star ($\bar{l}_\mathrm{massive}$).
The stars in the mass window are statistically significantly mass segregated if they fulfill
\<  \bar{\Lambda}_{(i)} ( \bar{m_{(i)} } ) - \frac{ \bar{\bar{\sigma}_{40}} } {\;\;\; \bar{l}_{(i),(i+40)} \;\;\; } > 1, \label{cond_seg} \>
where $\bar{\bar{\sigma}_{40}}$ is the standard deviation of $\bar{\bar{l}_{40}}$.

In order to be less influenced by outliers we propose an analogous measure, $\tilde{\Lambda}$, that uses the median MST edge length instead of the mean.
The equation for $\tilde{\Lambda}$ reads then
\< \tilde{\Lambda} ( \bar{ m_{(i)} } ) &=& \frac{ \bar{\tilde{l}_{40}} }{ \;\;\; \tilde{l}_{(i),(i+40)} \;\;\; }, \>
with $\tilde{l}_{(i),(i+40)}$ being the median edge length of the MST for the massive stars.
$\bar{\tilde{l}_{40}}$ is the sample average of the median edge length of $40$ random stars.
The condition for statistically significant mass segregation is analogous to eq. \ref{cond_seg}, using $\bar{\tilde{\sigma}_{40}}$, the standard deviation of $\bar{\tilde{l}_{40}}$.

\subsection{Inverse mass segregation indicated by $\bar{\Lambda}$}

\begin{figure}
\begin{center}
\includegraphics[width=0.4\textwidth]{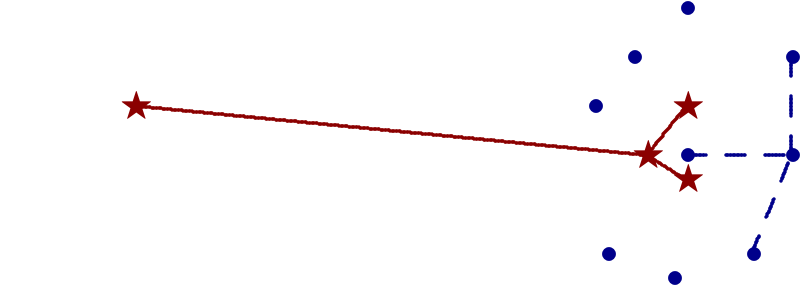}
\end{center}
\caption{\label{sketch_inverse}
Sketch illustrating the difficulty of MST methods in the presence of outliers, like Fig. \ref{sketch_substructure}, showing the locations of massive stars with star symbols and their MST as solid lines.
Here the MST for random stars (one example with dashed lines) only rarely contains the long distance to the outlying massive star.
Thus, the random MST edge length is shorter on average compared to the massive star MST edge length, so that $\bar{
\Lambda}$ gives the result ``inversely mass segregated'', although still most of the massive stars are in the centre of the cluster.
}
\end{figure}

At early times of the $1000\ \Msun$ simulation $\bar{\Lambda}$ is smaller than unity, suggesting
{\it inverse mass segregation}.
Here we show that this result is strongly driven by a single massive
outlier.
The situation  is illustrated schematically in Fig. \ref{sketch_inverse}, showing a mass segregated cluster with one massive outlier.
Within the cluster, the MST of the massive stars (solid lines) has short edges, shorter than the medium length edges of a random MST (dashed).
Without the isolated star, the $\Lambda$ method would demonstrate that
the cluster is mass segregated; however when the single  isolated massive star is included in the computation of $\bar{\Lambda}_\mathrm{massive}$, the system
would instead be classified as ``inversely mass segregated''.

\begin{figure}
\begin{center}
\includegraphics[width=0.4\textwidth]{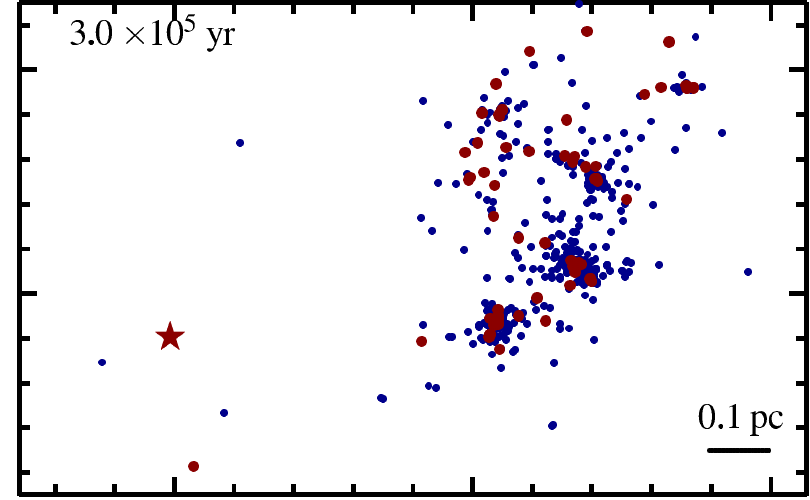}
\end{center}
\caption{\label{snapshot}
Projected positions of the sink particles in the $1000\ \Msun$ simulation at a time of 300 000 yr, when 413 sink particles have been formed ($m>0.08\ \Msun$).
Sink particles with a mass larger than 1 Msun are marked by the larger dots.
The star marks the position of a rather massive sink particle ($4.2\ \Msun$), that has formed earlier on at this fairly isolated location and continued to accrete.
}
\end{figure}

\begin{figure}
\begin{center}
\includegraphics[width=0.4\textwidth]{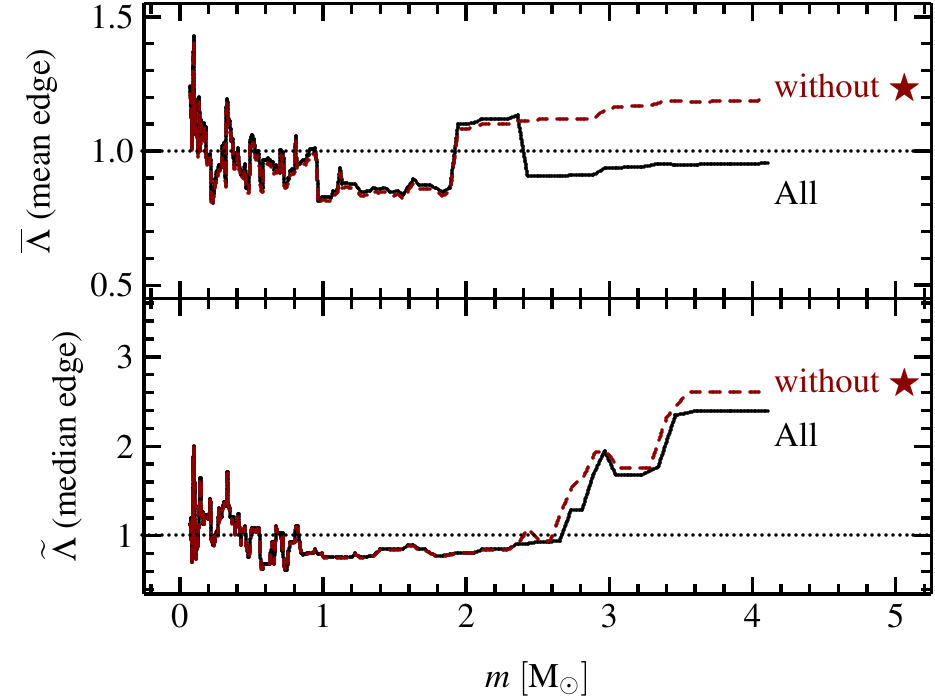}
\end{center}
\caption{\label{comparison}
$\bar{\Lambda}$ (top) and $\tilde{\Lambda}$  calculated for the snapshot shown in Fig. \ref{snapshot}, using all sink particles (solid lines)
and excluding the isolated massive sink marked by a star in Fig. \ref{snapshot}
(dashed lines).
Removing the edge to the isolated massive  sink changes the result of $\bar{\Lambda}$ from ``inversely mass segregated'' to ``mass segregated''.
$\tilde{\Lambda}$ is more robust to the removal of the isolated massive  sink and shows only minor changes.
Note that the absolute values of $\bar{\Lambda}$ and $\tilde{\Lambda}$ are not directly comparable.
}
\end{figure}

A concrete example from the $1000\ \Msun$ simulation is given in Fig. \ref{snapshot}, showing a snapshot of the sink particle distribution at an age of $3\times 10^{5}\ \mathrm{yr}$.
At this time 413  sink particles with masses $> 0.08\ \Msun$ have been formed.
Sinks with $m > 1\ \Msun$ are shown with larger dots, which are generally associated with subclusters.
The big star marks the position of the outlier, a sink that has at this time a mass of $4.2\ \Msun$ (the 16th most massive sink).
The results for $\bar{\Lambda}$ are given by the solid line in the upper panel of Fig. \ref{comparison}.
$\bar{\Lambda} < 1$ for $m > 1\ \Msun$, i.e. inverse mass segregation (although there is a ``bump'' around $2\ \Msun$).
The standard deviation of $\bar{\Lambda}$ is $0.16$, so that this result is not statistically significant.
But, when the outlier is removed, then $\bar{\Lambda} \approx 1.2 $  for $m > 2 \ \Msun$ (dashed line in the upper panel of Fig. \ref{comparison}), i.e. now the region is mass segregated.

The use of the median instead of the mean when calculating $\Lambda$ leads to more robust results with respect to outliers.
In the lower panel of Fig. \ref{comparison} $\tilde{\Lambda}$ is shown, either including the outlier (solid line) or omitting it (dashed line).
Both cases lead to the essentially identical curves, indicating that the sinks with $m > 2.5 \ \Msun$ are mass segregated.
(The standard deviation of $\tilde{\Lambda}$ is $\approx 0.4$).
It should be noted that the absolute values of $\bar{\Lambda}$ and $\tilde{\Lambda}$ are not directly comparable, i.e., omitting the outlier, $\bar{\Lambda}=1.2$ does {\it not} imply ``less'' mass segregation than $\tilde{\Lambda}=2$).

 The above results demonstrate the obvious point that a single measure cannot
contain all the information about the system - i.e. there is no value
of either $\bar{\Lambda}$ or $\tilde{\Lambda}$ that uniquely demonstrates
the actual situation shown in Figure 3 (i.e. a generally mass segregated
situation plus a single massive outlier). In the following section we
instead present plots of stellar mass versus surface density which
can be readily compared with observations.

\section{Measuring mass segregation with stellar surface densities}\label{sec_sigma}

In the previous Section we discussed the advantage of utilising the MST to quantify mass segregation in substructured regions, but also the fact that the
resulting statistic is very sensitive to outliers, especially
when the mean edge length is used.
Here we present as an alternative the use of a plot of stellar surface densities versus stellar mass, which can be also used in the presence of substructure.
It allows one to have an immediate grasp of the state of mass segregation
and the presence of outliers.

For the evaluation of the local surface number density around a star we follow the approach of \citet{vonhoerner1963} and \citet{casertano+hut1985}.
As we analyse the data in projection, we define for a star the local surface number density as
\< \Sigma &=& \frac{6-1}{\pi r_{6}^2}, \>
where $r_{6}$ is the distance to the sixth nearest neighbour of the star.
In the choice of the sixth nearest neighbour we follow \citet{casertano+hut1985}, who found that for a total number of particles between 30--1000 this choice is a good compromise between locality of the density estimate and the amount of low-number fluctuations.

\begin{figure}
\begin{center}
\includegraphics[width=0.4\textwidth]{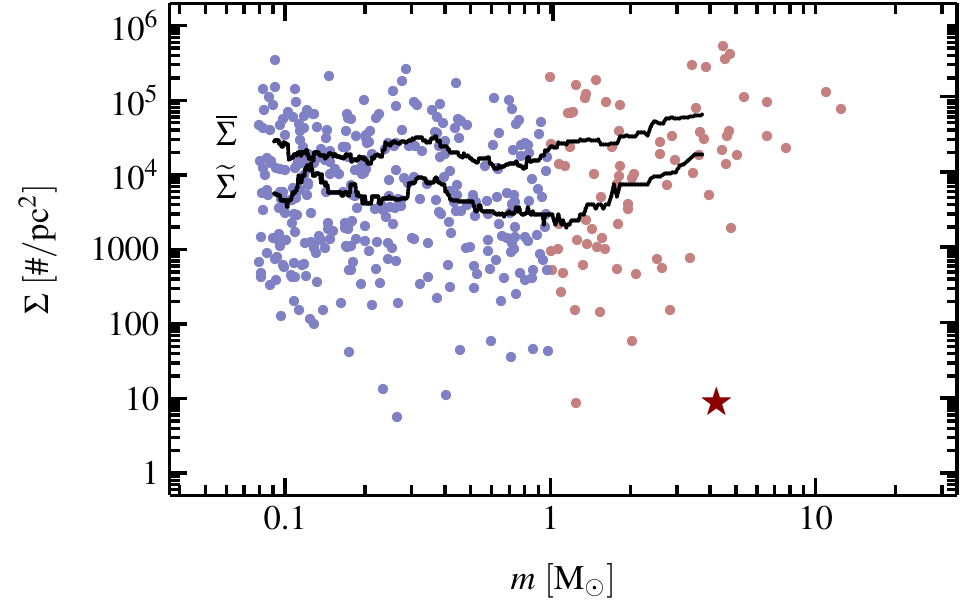}
\end{center}
\caption{\label{neighbour_densities}
Stellar surface density vs. mass, calculated for the snapshot shown in Fig. \ref{snapshot}.
Also shown is the moving window mean ($\bar{\Sigma}$) and median ($\tilde{\Sigma}$) surface density (window size 40 sinks).
The star marks the position of the outlier in Figure \ref{snapshot}.
}
\end{figure}

Figure \ref{neighbour_densities} gives an example of a plot of stellar density versus the sink particle mass for the snapshot of the $1000\ \Msun$ simulation shown in Figure \ref{snapshot}.
The dots show the local estimates for each sink particle, spanning five orders of magnitude.
It can immediately be seen that, whereas low mass sinks are found at all
surface densities, there are few massive  sinks in regions of low surface density. It is this relative deficit of points in the lower right of the plot
that drives the correlation between mass and surface density as is
evidenced by the rise in both median and mean mass with surface density at
masses $>1\ \Msun$. Note that the inclusion or omission of the
`isolated'' massive sink (marked with a star, as in Figure \ref{snapshot}) 
would have little effect on the median and mean trends. Thus the plot 
allows one to readily discern the outliers without allowing them
to dominate the inferences drawn about the distribution.
The differences in the surface densities of low- and high-mass sinks can be more formally quantified by a two-sample Kolmogorov-Smirnov test.
For this we divide the data shown in Fig. \ref{neighbour_densities} into a low-mass sample ($0.08\ \Msun < m \leq 1\ \Msun$, $n=334$) and a high-mass sample ($m > 1\ \Msun$, $n=79$).
The probability for those samples to obey the same distribution function is only 2.5\% ($\text{p-value} = 0.025$), i.e., the surface density distribution of the more massive sinks is more than $2\sigma$ deviating from the one of low-mass sinks.

\begin{figure}
\includegraphics[width=8.5cm]{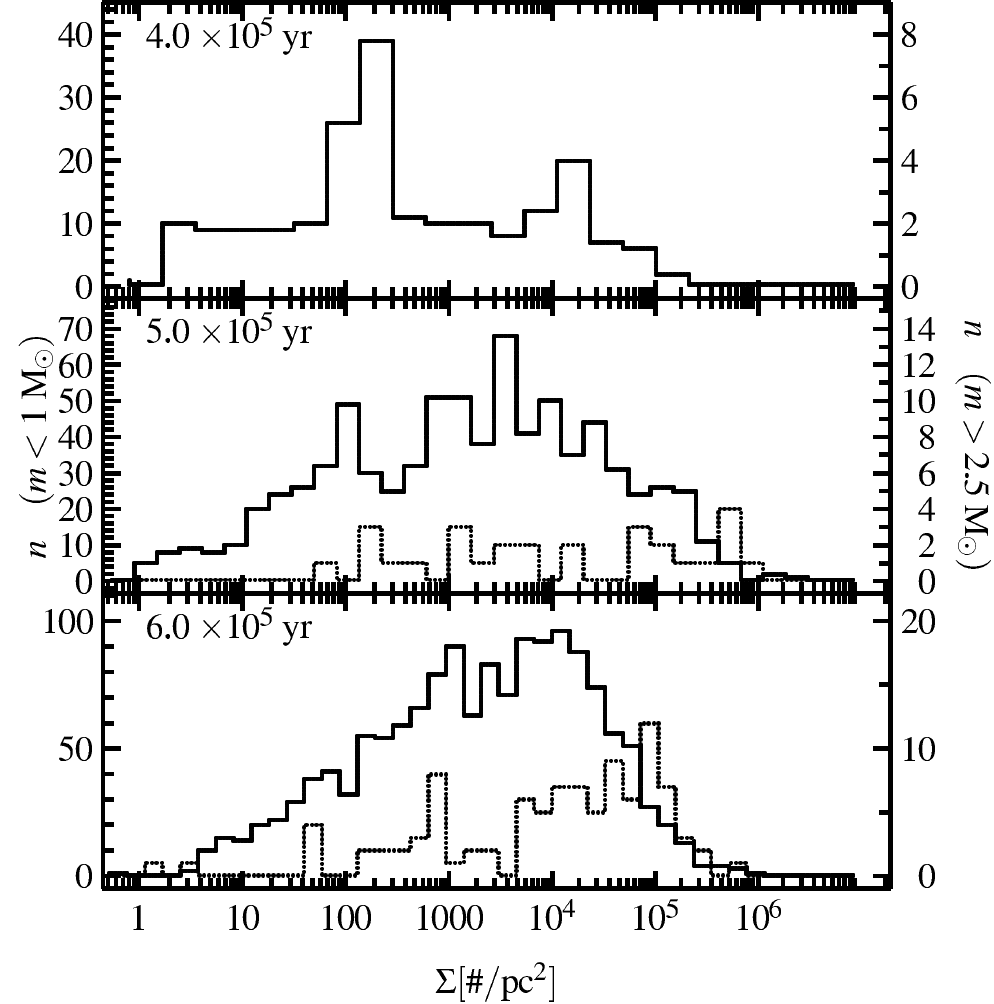}
\caption{\label{density_evolution}
Histograms of the surface density of sink particles of the $10^4\ \Msun$ simulation in the mass ranges $0.08$--$2.5\ \Msun$ (solid) and $>2.5\ \Msun$ (dotted, right $y$ axis scaled to 5 times the values for the lower range, as less sink particles have high masses).
The first sink particle formed at $3.2 \times 10^{5}\ \mathrm{yr}$.
The lower mass sink particles move to higher surface densities, as the subclusters contract, but mergers of subclusters mean that some sinks continue to be found at low  surface densities.
Massive sink particles have generally higher surface densities, although some can be found at low surface densities because they formed in isolation.
}
\end{figure}

Figure \ref{density_evolution} shows a  time sequence for the evolution of
the surface density as a function of stellar mass, in this case in
the case of the $10^4\ \Msun$ simulation which forms a number
of clusters (see Figure 2 of \citealp{maschberger-etal2010}  for the spatial distribution of the sinks). Evidently a large range of surface densities
is present in each snapshot, with the low density tail being contributed
both by regions of newly formed sinks together with clusters whose
density has been temporarily lowered during mergers.  The maximum surface density however
increases over the period $4 \times 10^5$--$6 \times 10^5$ yr as clusters merge and contract.
As  more massive sink particles ($>2.5\ \Msun$) are formed,
they tend to populate the higher end of the surface density distribution
(see dotted histograms in lower two panels).
The two-sampe KS probabilities are 0.8\% at $5 \times 10^5\ \mathrm{yr}$ and $6\times 10^{-9}$ at $6 \times 10^5\ \mathrm{yr}$, dividing the sample at $2.5\ \Msun$.

Note that even the more massive sink particles are occasionally found at low
surface densities, as evidenced by the tail in the bottom panel of Fig. \ref{density_evolution}.
For example, there are two massive sink particles ($m = 2.8\ \Msun$ and $m=2.6\ \Msun$) with very low surface densities at $6\times 10^{5}\ \mathrm{yr}$ (see lowest panel of Figure \ref{density_evolution}).
They have formed and grown in relative isolation (one in the unbound part and the other in an outward-moving filament of the bound part of the cloud) and at the moment of the snapshot no other sink particle has formed nearby.
In the time until the end of the simulation, some 50000 years later, a few sink particles form in the vicinity of both massive sinks, which themselves might in nature fragment to a binary or multiple system.

\section{Summary and discussion}\label{conclusions}

Hydrodynamical simulations of star forming regions \citep{bonnell-etal2003,bonnell-etal2008,maschberger-etal2010} show a general trend of mass segregation within the first few 100 000 years.
This follows with both using the minimum spanning tree technique ($\Lambda$, \citealp{allison-etal2009a}) and the distribution of local sink particle densities vs. mass.
We note that the MST technique using the mean edge length, as suggested by \citet{allison-etal2009a}, can give results that can be dominated by the presence of even a single star that in relative isolation from the main region.
We suggest that using the median MST edge length (instead of the mean) is more robust against such outliers.
The combined usage of mean and median indicates the trend for mass segregation of the majority of the massive stars as well as the presence of outliers.

 Alternatively, it proves useful to study mass segregation with  the $m$-$\Sigma$ plot (mass vs. stellar surface number density), which can be applied in substructured regions and makes the presence of massive outliers immediately perceptible.

The observational results for Taurus, showing inverse $\Lambda$ mass segregation  \citep{ParkerBouvierGoodwin-2011} and a concentration of massive stars at subcluster centres with some ``isolated'' massive stars \citep{KirkMyers-2011} {\it may} be compatible with the simulations.
Although Taurus is a much sparser system than the one modelled here,
it would be interesting to see whether the distribution of sources
in the $m$-$\Sigma$ plane is qualitatively similar to the simulations

In the simulations, ``massive'' sink particles ($m>2.5\ \Msun$) can form and stay for extended periods in relative isolation (stellar surface density $< 10/\mathrm{pc}^2$, i.e. $< 10^{-4}$
of the median surface density in the simulation).
However, they are a small minority of only up to $\approx$ 2--3\% of all massive sinks (1 of 33 in the 1000\ \Msun\  calculation at $t=3\times 10^5\ \mathrm{yr}$, 2 of 98 in the $10^4\ \Msun$ calculation, $t=6\times 10^5\ \mathrm{yr}$).

The softened gravitational potential hinders strong dynamical interactions of sink particles in the calculations and so also ejections of massive sinks.
Therefore the occurrence of isolated massive sinks via the dynamical channel is possibly underestimated.
 Thus the figure of 2--3\% represents the fraction of massive sinks that are relatively isolated and have formed in situ in the simulations. 
 We cannot at present quantify the numbers of ejected massive sinks that might additionally appear in isolation.

\section{Acknowledgements}
Th. M. acknowledges funding through {\sc constellation}, an European Commission FP6 Marie Curie Research Training Network, and  the Stellar Populations and Dynamics Research Group at the Argelander-Institut f{\"ur} Astronomie.
We would like to thank Pavel Kroupa and Andreas K{\"u}pper for helpful comments regarding the manuscript.

\bibliographystyle{mn_mod}
\bibliography{refs_tm_wrk}
\label{lastpage}
\bsp

\end{document}